\pdfoutput=1
\documentclass[conference,transmag]{IEEEtran}

\usepackage{cite}

\ifCLASSINFOpdf
  \usepackage[pdftex]{graphicx}
    \graphicspath{{./figs/}}
      \DeclareGraphicsExtensions{.pdf,.jpg,.png}
\else
        \usepackage[dvips]{graphicx}
    \graphicspath{{./figs/}}
      \DeclareGraphicsExtensions{.eps}
\fi

\ifCLASSOPTIONcompsoc
 \usepackage[caption=false,font=normalsize,labelfont=sf,textfont=sf]{subfig}
\else
 \usepackage[caption=false,font=footnotesize]{subfig}
\fi

\usepackage[utf8]{inputenc}
\usepackage[T1]{fontenc}
\usepackage{siunitx}
\usepackage{graphicx}
\usepackage{mathtools}
\usepackage{tikz}
\usepackage{circuitikz}
\usetikzlibrary{arrows,matrix,positioning,plotmarks,shapes.geometric,shapes,shapes.misc,shapes.multipart,calc}
\usepackage{pgfplots}
\pgfplotsset{compat=newest}
\usepackage{epstopdf}

\usepackage[disable]{todonotes}
\usepgfplotslibrary{external}
\usepackage{eso-pic}

\catcode`@=11
\def\frownfill{$\scriptscriptstyle\m@th\mathord\frown$}
\def\bow#1{\vbox{\m@th\ialign{##\crcr
      \hfil\frownfill\hfil\crcr\noalign{\kern-0.2\p@\nointerlineskip}
      $\hfil\displaystyle{#1}\hfil$\crcr}}}
\def\bbow#1{\vbox{\m@th\ialign{##\crcr
     \hfil\frownfill\hfil\crcr\noalign{\kern-0.7\p@\nointerlineskip}
     \hfil\frownfill\hfil\crcr\noalign{\kern-0.3\p@\nointerlineskip}
      $\hfil\displaystyle{#1}\hfil$\crcr}}}
\def\widefrownfill{$\m@th\mathord\frown$}
\def\widebow#1{\vbox{\m@th\ialign{##\crcr
      \hfil\widefrownfill\hfil\crcr\noalign{\kern-0.9\p@\nointerlineskip}
      $\hfil\displaystyle{#1}\hfil$\crcr}}}
\def\widebbow#1{\vbox{\m@th\ialign{##\crcr
     \hfil\widefrownfill\hfil\crcr\noalign{\kern-1.8\p@\nointerlineskip}
     \hfil\widefrownfill\hfil\crcr\noalign{\kern-0.9\p@\nointerlineskip}
      $\hfil\displaystyle{#1}\hfil$\crcr}}}

\AddToShipoutPicture*{\footnotesize\sffamily\raisebox{1cm}{\hspace{1.65cm}\fbox{\parbox{\textwidth}{\copyright~2016 IEEE. Personal use of this material is permitted. Permission from IEEE must be obtained for all other uses, in any current or future media, including reprinting/republishing this material for advertising or promotional purposes, creating new collective works, for resale or redistribution to servers or lists, or reuse of any copyrighted component of this work in other works.}}}}
\def\defineCMYKcolor(#1,#2,#3,#4)#5{%
    \pgfmathsetmacro{\myc}{#1/255}%
    \pgfmathsetmacro{\mym}{#2/255}%
    \pgfmathsetmacro{\myy}{#3/255}%
    \pgfmathsetmacro{\myk}{#4/255}%
    \definecolor{#5}{cmyk}{\myc,\mym,\myy,\myk}%
}

\newcommand\blfootnote[1]{	\begingroup
	\renewcommand\thefootnote{}\footnote{#1}	\addtocounter{footnote}{-1}	\endgroup
}

\newcommand*{\TitleFont}{      \usefont{\encodingdefault}{\rmdefault}{m}{n}      \fontsize{24pt}{29pt}      \selectfont}

\begin{document}
\title{\TitleFont Electrothermal Simulation of Bonding Wire Degradation under Uncertain Geometries}

\author{\IEEEauthorblockN{Thorben~Casper\IEEEauthorrefmark{1,2}, Herbert~De~Gersem\IEEEauthorrefmark{2}, Renaud~Gillon\IEEEauthorrefmark{3}, Tomas~Gotthans\IEEEauthorrefmark{4},\\Tomas~Kratochvil\IEEEauthorrefmark{4}, Peter~Meuris\IEEEauthorrefmark{5}, Sebastian~Schöps\IEEEauthorrefmark{1,2}}\\[-0.5em]
\IEEEauthorblockA{\IEEEauthorrefmark{1}Graudate School of Computational Engineering, Technische Universität Darmstadt, 64293 Darmstadt, Germany}
\IEEEauthorblockA{\IEEEauthorrefmark{2}Institut für Theorie Elektromagnetischer Felder, Technische Universität Darmstadt, 64289 Darmstadt, Germany}
\IEEEauthorblockA{\IEEEauthorrefmark{3}ON Semiconductor, 9700 Oudenaarde, Belgium}
\IEEEauthorblockA{\IEEEauthorrefmark{4}Department of Radio Electronics, Brno University of Technology, 60190 Brno, Czech Republic}
\IEEEauthorblockA{\IEEEauthorrefmark{5}Magwel, 3000 Leuven, Belgium}\\[-0.5em]
}

\maketitle

\begin{abstract}

In this paper, electrothermal field phenomena in electronic components are considered. This coupling is tackled by multiphysical field simulations using the Finite Integration Technique~(FIT). In particular, the design of bonding wires with respect to thermal degradation is investigated. Instead of resolving the wires by the computational grid, lumped element representations are introduced as point-to-point connections in the spatially distributed model. Fabrication tolerances lead to uncertainties of the wires' parameters and influence the operation and reliability of the final product. Based on geometric measurements, the resulting variability of the wire temperatures is determined using the stochastic electrothermal field-circuit model.

\end{abstract}

\IEEEpeerreviewmaketitle

\blfootnote{This work is supported by the European Union within FP7-ICT-2013 in the context of the \emph{Nano-electronic COupled Problems Solutions} (nanoCOPS) project (grant no. 619166), by the \emph{Excellence Initiative} of the German Federal and State Governments and by the Graduate School of CE at TU Darmstadt.}

\section{Introduction}
\label{sec:introduction}

Micro- and nanoelectronic components are designed using computer simulations. Especially due to the continuous shrinking of elements, power densities increase and therefore thermal considerations in an early design stage are of major importance. This indicates the need for coupled electrothermal simulations. Additionally, small feature sizes lead to significant fabrication tolerances that need to be tackled by Uncertainty Quantification (UQ).

When designing bonding wires for the packaging of Integrated Circuits (ICs), the designer is left with the choice of its material and its thickness. There is a tradeoff between minimal cost and maximum performance. Moreover, the thinner the wire, the higher the probability of failure during operation. On the other hand, the length of a wire is predetermined by the geometry of the given package. While the material is commonly chosen according to economic aspects and its physical properties, the leftover design parameter is the wire's thickness. Bonding wire calculators allow to estimate appropriate parameters by simulation. Many electrothermal models have been proposed for dedicated bonding wire simulation. In particular, there are phenomenological models determined from measurement data and models derived analytically or by discretization from the electrothermal problem and combinations of those approaches, see e.g.~\cite{Nobauer_2000aa, Mouthaan_2001aa,Schuster_2000aa, Duque_2015aa} and the references therein. 

To incorporate all physical effects, field simulation of integrated and discrete semiconductor power devices is well established. It is typically based on volumetric space discretization using for example the Finite Element Method (FEM) or the Finite Integration Technique~(FIT)~\cite{Weiland_1977aa,Weiland_1996aa}. However, the treatment of dynamic electrothermal effects is still challenging due to the coupling~\cite{Alotto_2010aa} and in particular because of multirate and multiscale effects~\cite{Clemens_2001ac,Schoenmaker_2013aa}. Resolving small features as thin wires is such a multiscale problem and therefore, many commercial simulators include various surrogate models to avoid discretizing the bonding wire in the computational grid, e.g.~\cite{Mouthaan_2001aa}. 

In this paper, we discuss a framework for embedding lumped electrothermal bonding wire models into electrothermal field simulators. A nonlinear electric and thermal network based model is proposed and consistently coupled to the spatial discretization. As an application example, the global sensitivity of the bonding wires' temperatures w.r.t. their geometric parameters is investigated. This is necessary because manufacturing tolerances, measurement inaccuracies and model imperfections lead to deviations between simulation and reality. 

The paper is structured as follows, section \ref{sec:electrothermal_coupling} introduces the coupling between electromagnetics and heat in the continuous setting while section \ref{sec:discretization} introduces the used discretization approach. Section \ref{sec:uncertainty_quantification} explains the presence and treatment of uncertainties in the bonding wires. Results of the simulation are then discussed in section \ref{sec:simulation_results} while section \ref{sec:conclusions} concludes the paper.

 \section{Electrothermal Field Problem}
\label{sec:electrothermal_coupling}

On the one hand, if an electrical current is applied to a bonding wire, the temperature of the wire increases due to the Joule heating effect. On the other hand, a change in temperature of the wire leads to a change of the material parameters. Neglecting the temperature dependence of the volumetric heat capacity, the nonlinearity of electrical and thermal conductivities in temperature remains. After the introduction of the electrical and thermal sub-problems, this two-directional electrothermal coupling is described.

\subsection{Electrical Sub-Problem}

The distribution of electrical quantities can be described by the current continuity equation. Neglecting capacitive effects, only the stationary current problem
\begin{equation*}
- \nabla\cdot\sigma(T)\nabla\varphi = 0 
\end{equation*}
with suitable boundary conditions is considered. Here, $\sigma$ is the electrical conductivity, $\varphi$ the electrical potential and ${T = T(t)}$ the time dependent temperature. The spatial dependencies are suppressed to keep the notation short. A generalization to electroquasistatics is straightforward.

\subsection{Thermal Sub-Problem}

Thermal heat is distributed due to conduction, convection and radiation. In the general form, the transient heat equation describes conduction and is given by
\begin{equation*}
 \rho c\dot{T} - \nabla\cdot\lambda(T)\nabla T = Q(T,\varphi),
\end{equation*}
where $\rho c$ is the volumetric heat capacity and $\lambda$ the thermal conductivity. The power density $Q$ represents heat sources that affect the system. In this paper, we assume three different sources to contribute to this heat source
\begin{equation*}
 {Q(T,\varphi)=Q_{\text{el}}(T,\varphi)+Q_{\text{bnd}}(T)}+Q_{\text{bw}}(T,\varphi).
\end{equation*}
First, heat can be generated by the Joule heating term $Q_{\text{el}}$ resulting from the electrical contribution. Secondly, heat exchange with the environment is described by the boundary term $Q_{\text{bnd}}$. Thirdly, the considered bonding wire acts as an external heat source $Q_{\text{bw}}$ since it is not resolved by the grid (see section~\ref{sec:bonding_wire_model}). These quantities will be explained throughout this paper. The heat exchange with the environment is modeled as Dirichlet, adiabatic, convective or radiative conditions.

The boundary term $Q_{\text{bnd}} = Q_{\text{conv}} + Q_{\text{rad}}$ contains a contribution of convective and radiative effects given by
\begin{equation*}
 Q_{\text{conv}} = -\frac{1}{|V|}\int_{\partial V} \vec{q}_\text{\text{conv}}\cdot\,\text{d}\vec{A}, \quad Q_{\text{rad}} = -\frac{1}{|V|}\int_{\partial V} \vec{q}_\text{\text{rad}}\cdot\,\text{d}\vec{A}.
\end{equation*}
In these equations, $\vec{q}_\text{\text{conv}}$ and $\vec{q}_\text{\text{rad}}$ are the heat fluxes that leave a volume $V$ due to convection or radiation as given by
\begin{equation*}
	\vec{q}_\text{\text{conv}} = h\left[T_{\text{bnd}}(t)-T_\infty\right]\vec{n},\quad
 \vec{q}_\text{\text{rad}} = \varepsilon\sigma_{\text{SB}}\left[T_{\text{bnd}}^4(t)-T_\infty^4\right]\vec{n},
\end{equation*}
respectively. Here, $\vec{n}$ is the outward-pointing normal, $h$ the heat transfer coefficient, $\varepsilon$ the emissivity and $\sigma_{SB}$ the Stefan-Boltzmann constant. As boundary effects are dominated by the boundary nodes and the environment, $T_{\text{bnd}}$ is the temperature at the boundary and $T_\infty$ the ambient temperature.

\subsection{Electrothermal Problem}

By combining the transient heat equation with the stationary current problem, we obtain the nonlinear electrothermal system
\begin{align}
	- \nabla\cdot\sigma(T)\nabla\varphi &= 0, \label{eq:electrostatics_continuous} \\
	\rho c\dot{T} - \nabla\cdot\lambda(T)\nabla T &= Q(T,\varphi) \label{eq:heat_equation_continuous}
\end{align}
with suitable boundary and initial conditions. The Joule heating due to the stationary current problem is described by ${Q_{\text{el}} = (\nabla\varphi)^{\top}\sigma(T)\nabla\varphi}$ resulting from the electrothermal coupling from electrical to thermal side. The two-directional coupling is established by the temperature dependence of $\lambda$ and $\sigma$.

\todo[inline]{work on this paragraph}

 \section{Discretization by FIT}
\label{sec:discretization}

\subsection{Field Model}

\newcommand{\ve}{\protect\bow{\rm\bf e}}
\newcommand{\vt}{\protect\bow{\rm\bf t}}
\newcommand{\fj}{\protect\bbow{\rm\bf j}}
\newcommand{\fq}{\protect\bbow{\rm\bf q}}
\newcommand{\mbf}{\mathbf}

\newcommand{\G}{\mathbf{G}}
\newcommand{\Ss}{\widetilde{\mathbf{S}}^{\top}}
\newcommand{\fMsigma}{\mathbf{M}_{\sigma}}
\newcommand{\fMlambda}{\mathbf{M}_{\lambda}}
\newcommand{\fMrhoc}{\mathbf{M}_{\rho c}}

The coupled electrothermal problem is discretized in space using the FIT \cite{Weiland_1977aa,Weiland_1996aa} on a staggered 3D hexahedral grid pair. For simplicity of notation, a staircase material approximation at the primary grid is assumed, i.e., each primary cell is assumed to consist of a homogeneous material. The discrete unknowns, i.e., the electric potentials $\Phi$ as well as the temperatures $\mathbf{T}$ are allocated at the nodes of the primary grid. The voltages and the temperature drops at the primary edges are found as differences, i.e., $\ve=-\G\Phi$ and $\vt=-\G\mbf{T}$ where $\G$ is the discrete gradient operator consisting of $0$, $1$ and $-1$ entries according to the topology of the primary grid. The currents $\fj$ and the heat fluxes $\fq$ are allocated at the facets of the dual grid. The currents and heat fluxes accumulating at the dual cells are calculated by $\widetilde{\mathbf{S}}\fj$ and $\widetilde{\mathbf{S}}\fq$ where $\widetilde{\mathbf{S}}$ is the discrete divergence operator containing $0$, $1$ and $-1$ entries according to the topology of the dual grid. The duality of the grids gives rise to the property $\G=-\Ss$.
\todo[inline]{name a quantity for $\widetilde{\mathbf{S}}\fj$ and $divdfit\fq$}

The currents~$\fj=\fMsigma\ve$ and the heat fluxes~$\fq=\fMlambda\vt$ are related to the voltages and temperature drops by the electrical and thermal conductance matrices~$\fMsigma$ and ~$\fMlambda$, respectively. In the case of a mutually orthogonal grid pair, every primary edge crosses a unique dual facet perpendicularly. In that case, the primary edges and dual facets can be indexed similarly and the material matrices $\fMsigma$ and $\fMlambda$ are diagonal with the entries
\begin{align*}
	\mbf{M}_{\sigma,i,i} = \frac{\overline{\sigma}_i\tilde{A}_i}{\ell_i} 
	\quad
	\text{and}
	\quad
	\mbf{M}_{\lambda,i,i} = \frac{\overline{\lambda}_i\tilde{A}_i}{\ell_i}
\end{align*}
where $\ell_i$ is the length of primary edge $i$ and $\tilde{A}_i$ is the area of dual facet $i$. The material parameters $\overline{\sigma}$ and $\overline{\lambda}$ are found by a volumetric averaging of the corresponding parameters of the primary cells touching the considered primary edge.

The thermal capacitance matrix $\fMrhoc$ relates the heat power to the temperature change, i.e., $\mathbf{Q}_{\dot{\mathbf{T}}}=\fMrhoc\dot{\mbf{T}}$ and operates between the primary nodes and the dual cells. Also here, a one-to-one relation is present and the indexing scheme is shared. The diagonal entries of $\fMrhoc$ read
\begin{equation*}
 \mbf{M}_{\rho c,j,j} ={\overline{\rho c}_j\tilde{V}_j}
\end{equation*}
where $\tilde{V}_j$ is the volume of dual cell $j$. Here, $\overline{\rho c}_j$ is obtained by averaging the volumetric heat capacity of the primary cells touching the considered dual cell $j$.

The heat generated by the current is calculated at the dual cells, i.e., the voltages $\ve$ are interpolated to the midpoints of the primary cells yielding $\vec{E}_k$ where $k$ counts over all primary cells. There, the power density is calculated by~${Q_{{\rm el},k}=\sigma_k\vec{E}_k\cdot\vec{E}_k}$. The electrical contribution to the power then follows from
\begin{equation*}
	\mathbf{Q}_{\rm el}(T,\varphi) = \tilde{V}_j \overline{\mathbf{q}}_j
\end{equation*}
where $\overline{\mathbf{q}}_j$ results from averaging the powers from the primary cells to the primary nodes.

The topological operators $\G$ and $\Ss$ and the material matrices $\fMsigma$, $\fMlambda$ and $\fMrhoc$ are put together in the discrete counterpart to \eqref{eq:electrostatics_continuous} and \eqref{eq:heat_equation_continuous}, i.e.,
\begin{align}
	-\widetilde{\mathbf{S}}\mathbf{M}_{\sigma}(\mathbf{T})\mathbf{G}\mathbf{\Phi} &= \mathbf{0}, \label{eq:electrostatics_discrete} \\
	\mathbf{M}_{\rho c} \dot{\mathbf{T}} - \widetilde{\mathbf{S}}\mathbf{M}_{\lambda}(\mathbf{T})\mathbf{G} \mathbf{T} &= \mathbf{Q}(\mathbf{T},\mathbf{\Phi}). \label{eq:heat_equation_discrete}
\end{align}
The degrees of freedom are the discrete temperature vector $\mathbf{T}=\mathbf{T}(t)$ and the electrical potential vector $\mathbf{\Phi}=\mathbf{\Phi}(t)$, while ${\mathbf{Q}=\mathbf{Q_{\text{el}}} + \mathbf{Q_{\text{bnd}}} + \mathbf{Q_{\text{bw}}}}$ is the discrete representation of $Q$. It includes Joule heating by the field model, the boundary term for convective and radiative effects as well as the self-heating of the bonding wires as explained in the next section. Subsequently, the time is discretized by the implicit Euler method.

\begin{figure}[b]
 \centering
 \resizebox{\linewidth}{!}{
\includegraphics{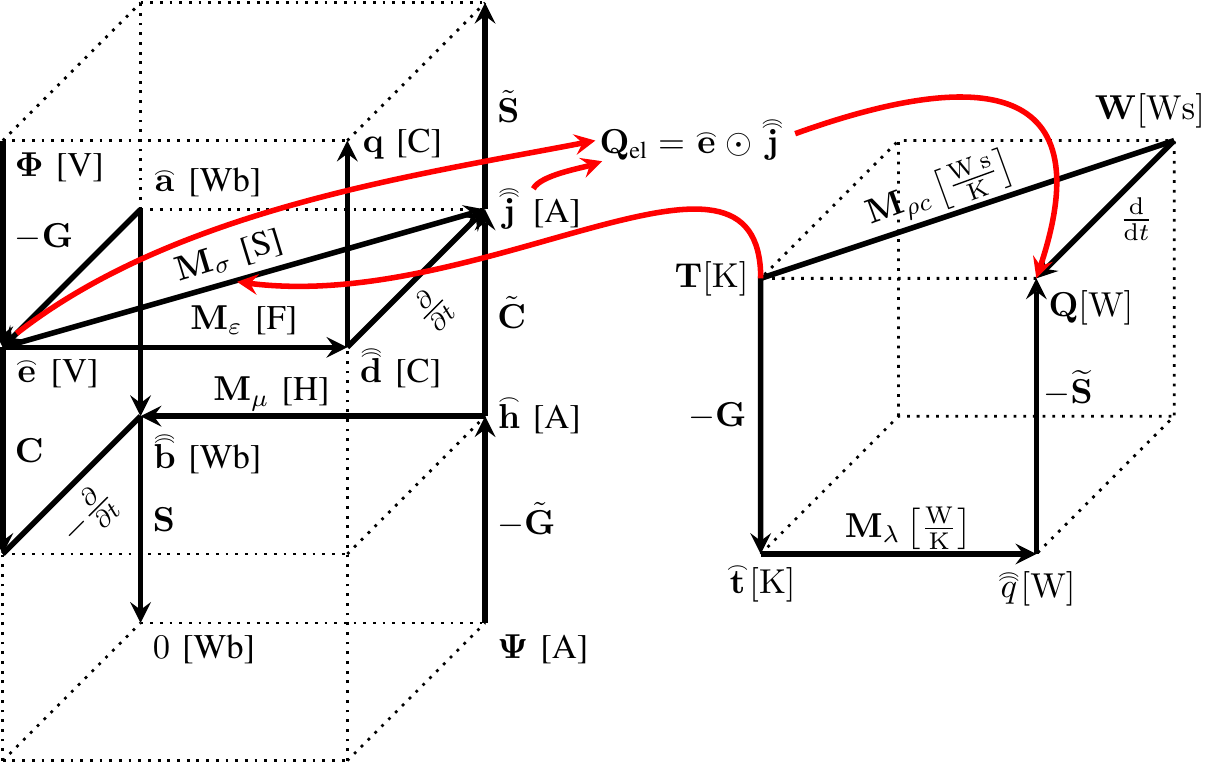}

}  \caption{Discrete electrothermal house.}
 \label{fig:electrothermal_house_discrete}
\end{figure}

For an overview of the involved quantities and their relation, the discrete electrothermal house based on \cite{Clemens_2005aa} and \cite{Alotto_2010aa} is shown in Fig.~\ref{fig:electrothermal_house_discrete}. The figure consists of two parts showing the Maxwell house on the left hand side and the thermal house on the right hand side. The coupling is established due to the Joule heating term and the nonlinear electrical conductivity as illustrated.

\subsection{Bonding Wire Model}
\label{sec:bonding_wire_model}

\begin{figure}[b]
 \centering
 \includegraphics{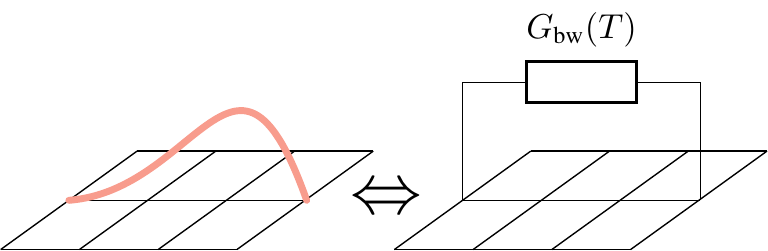}

  \caption{Bonding wire modeling by a lumped element approach.}
 \label{fig:bw_inclusion_mesh}
\end{figure}

To account for the different scales of the bonding wires in comparison to any other microelectronic components in their vicinity, the wires are not resolved by the grid but instead modeled by a lumped element approach. Assuming that conduction is dominant compared to capacitive effects, pairs of mesh points are connected by an electrothermal conductance only. Here, the conductance~$G_{\rm bw}$ serves as a placeholder for both the electrical and thermal conductance~$G_{\text{bw}}^{\text{el}}$ and $G_{\text{bw}}^{\text{th}}$. This approach is illustrated in Fig.~\ref{fig:bw_inclusion_mesh}.

For the implementation, the conductance matrix 
\begin{equation*}
 \mathbf{G}_{\text{bw}} =
 \begin{bmatrix}
  G_{\text{bw}} & -G_{\text{bw}} \\
  -G_{\text{bw}} & G_{\text{bw}}
 \end{bmatrix}
\end{equation*}
needs to be stamped to the correct positions in the system matrices of \eqref{eq:electrostatics_discrete} and \eqref{eq:heat_equation_discrete}. If this is done for all $N_\mathrm{bw}$ bonding wires present in the model, the extended electrothermal system reads
\begin{align*}
 \widetilde{\mathbf{S}}\mathbf{M}_{\sigma}(\mathbf{T})\widetilde{\mathbf{S}}^{\top}\mathbf{\Phi}+\sum_{j=1}^{N_\mathrm{bw}} \mathbf{P}_{j}G_{\text{bw},j}^{\text{el}}(T_{\text{bw},j})\mathbf{P}_{j}^{\top}\mathbf{\Phi} &= \mathbf{0}, \\
 \mathbf{M}_{\rho c} \dot{\mathbf{T}} + \widetilde{\mathbf{S}}\mathbf{M}_{\lambda}(\mathbf{T})\widetilde{\mathbf{S}}^{\top} \mathbf{T} + \sum_{j=1}^{N_\mathrm{bw}} \mathbf{P}_{j}G_{\text{bw},j}^{\text{th}}(T_{\text{bw},j})\mathbf{P}_{j}^{\top}{\mathbf{T}} &= \mathbf{Q}(\mathbf{T}).
\end{align*}
Here, $\widetilde{\mathbf{S}}^{\top} = -\mathbf{G}$ represents the negative gradient matrix and $\mathbf{P}_{j}$ is a bonding wire gradient vector consisting of $0, 1\text{ and } -1$ entries that additionally handles the incidences between the contacts of the bonding wire $j$ and the grid. The above equation assumes a linear temperature distribution across each lumped element with its average defined by 
\begin{equation}
T_{\text{bw},j}=\mathbf{X}^{\top}_j \mathbf{T}, \label{eq:wire_temperature}
\end{equation}
The vector $\mathbf{X}_j$ contains two $1/2$ entries and averages the temperature from both bonding wire connection nodes. To account for nonlinear temperature distributions, a single bonding wire can be modeled by a more complex model or by a number of concatenated lumped elements resulting in a piecewise linear temperature distribution.

With these quantities defined, the Joule heating of a single bonding wire reads
\begin{align*}
Q_{\text{bw},j} = \mathbf{\Phi}^{\top}\mathbf{P}_{j}G_{\text{bw},j}^{\text{el}}(T_{\text{bw},j})\mathbf{P}_{j}^{\top}\mathbf{\Phi}.
\end{align*}
Finally, the contribution of all bonding wires to the right hand side of \eqref{eq:heat_equation_discrete} is given by
\begin{equation*}
 \mathbf{Q}_{\text{bw}} = \sum_{j=1}^{N_\mathrm{bw}}\mathbf{X}_j Q_{\text{bw,j}}.
\end{equation*}

 \section{UQ for Bonding Wires}
\label{sec:uncertainty_quantification}

While the methodology has been presented in the previous sections, an application example is given in the following. Bonding wires attached to a chip are modeled using the lumped element approach to realize an electrothermal simulation of the full package. Additionally, UQ is applied to account for the variations in the bonding wire lengths. 

\subsection{Exemplary Package}

\begin{figure}[b]
 \centering
 \subfloat[Top view\label{fig:X-ray_topview}]{\includegraphics[height=12em]{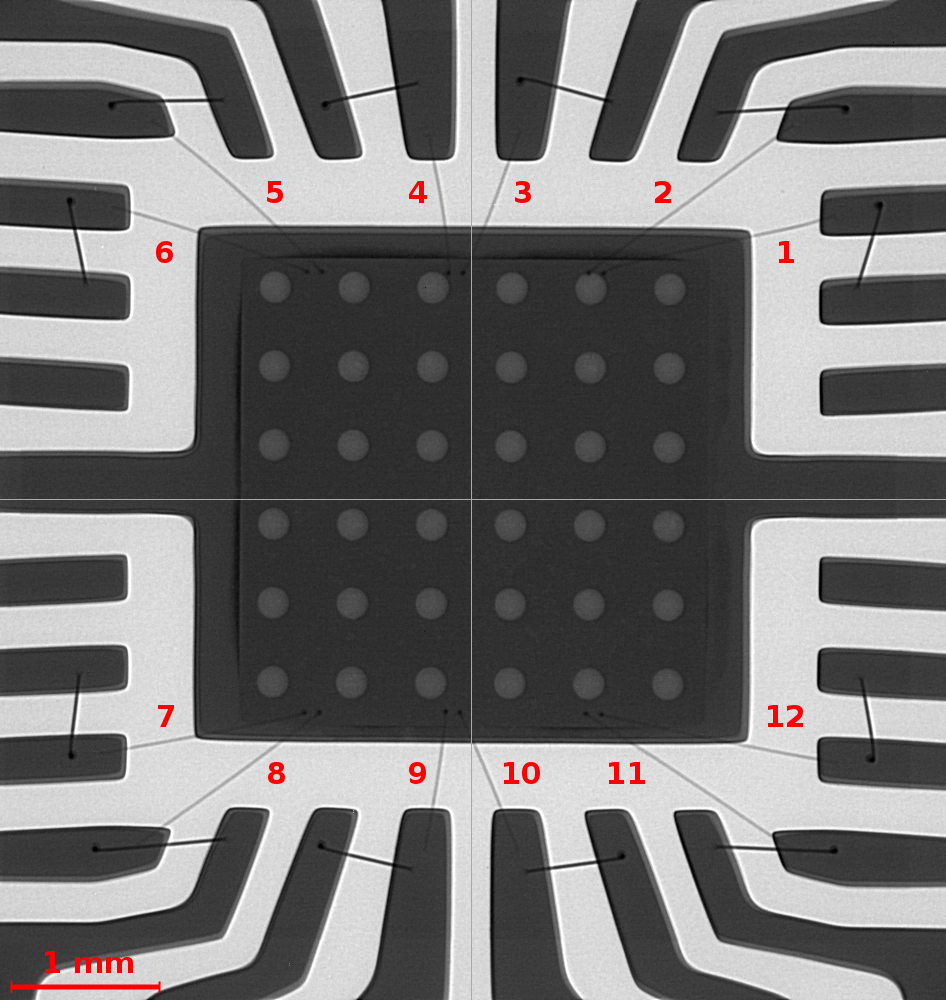}}\quad
 \subfloat[Perspective view\label{fig:X-ray_sideview}]{\includegraphics[height=12em]{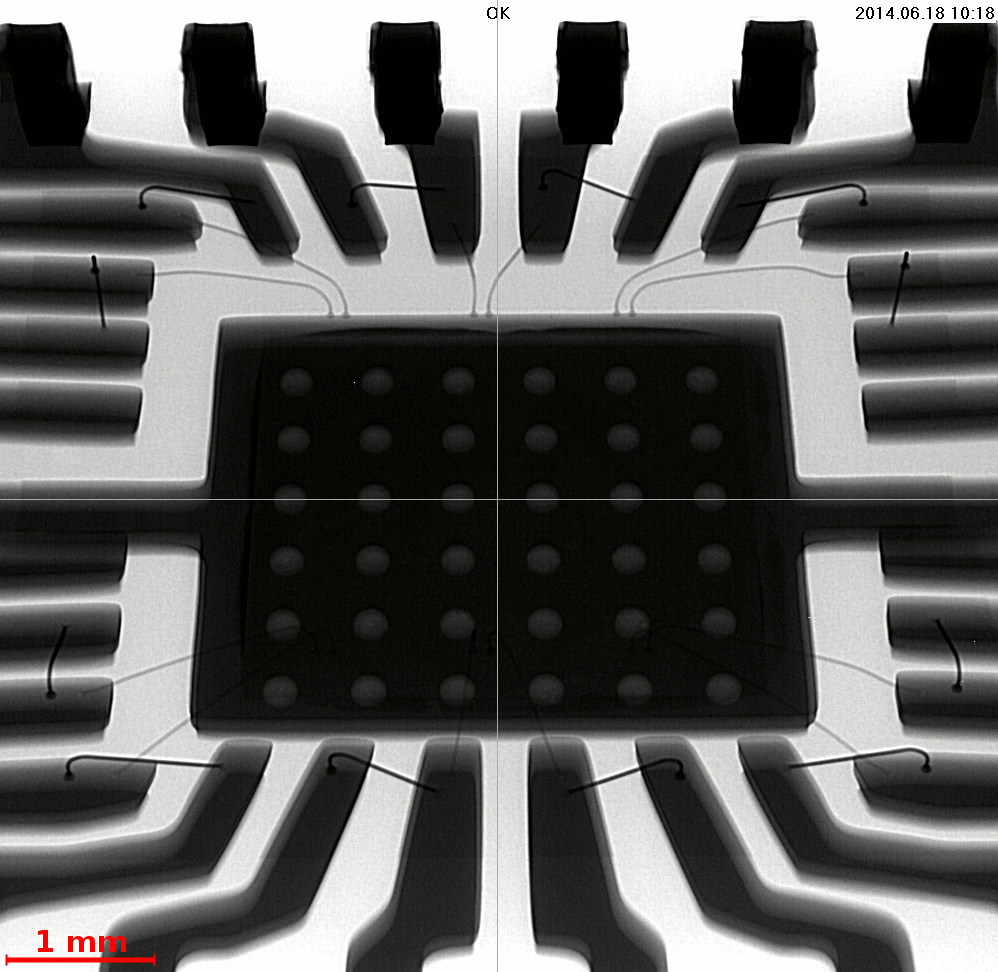}}
 \caption{X-ray photos of the investigated chip.}
 \label{fig:X-ray}
\end{figure}

The example features $28$ contacts and ${N_\mathrm{bw}=12}$ bonding wires as shown in the X-ray pictures in Fig.~\ref{fig:X-ray}. Each wire connects the chip with one of the contact pads. Because only the contacts are accessible from the outside, a constant voltage is always applied over two adjacent bonding wires (e.g. wires $3$ and $4$ or wires $7$ and $8$).

The time until the wire fails depends on the applied voltage, the material properties and the geometry of the wire. Assuming that we know the voltage and the conductivity of the material accurately, the uncertainty is only related to the wires' geometry. While the thickness of the wires can be fabricated very accurately, the only unknown parameter remains to be the length of the wire. This length is not a priori known as it highly depends on the bonding process. By using the X-ray pictures in Fig.~\ref{fig:X-ray}\subref{fig:X-ray_topview} and \subref{fig:X-ray_sideview}, the lengths have been measured after fabrication.

\subsection{Modeling of the Uncertain Bonding Wire Length}
\label{sec:uncertain_geometry}

\begin{figure}[t]
 \centering
 \subfloat[Exact position on contact pad.\label{fig:bwire_position_perfect}]{
\includegraphics{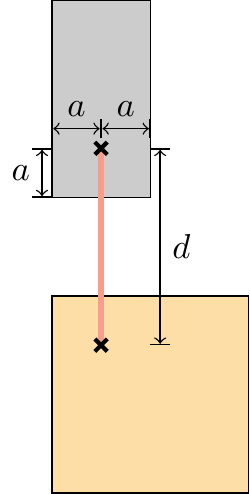}
 }\qquad
 \subfloat[Elongation due to misplacement.\label{fig:bwire_position_variable}]{\includegraphics{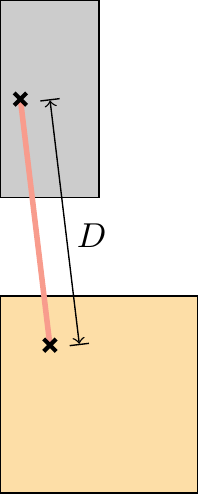}
 }\\ \vspace{-1em}
 \subfloat[Elongation due to bending.\label{fig:pad2chip_bw_uncertain}]{\includegraphics{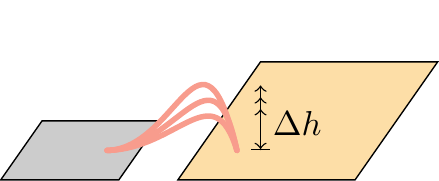}

 }
 \caption{Variability of the bonding wire length due to construction tolerances.}
\end{figure}

The correct length of a bonding wire depends on three parameters. First, the minimal length of a wire is given by the direct distance $d$ between contact pad and chip as shown in Fig.~\ref{fig:bwire_position_perfect}. Here, it has been assumed that the bonding was done such that the position of the wire's end points is exactly as planned by the designer. This means that the connection point on the contact pad is equally spaced (length $a$) to its edges. Secondly, any deviation from the perfect position on the contact pad leads to an elongation $\Delta s$ that adds up to the corrected distance $D = d + \Delta s$ as depicted in Fig.~\ref{fig:bwire_position_variable}. Thirdly, any additional bending results in an additional elongation $\Delta h$ (see Fig.~\ref{fig:pad2chip_bw_uncertain}) giving the total wire length $L=d +\Delta s + \Delta h$. Due to the camera angle in Fig.~\ref{fig:X-ray_sideview}, the elongation $\Delta h$ could only be determined for $6$ wires. For the other wires, the average value of these $6$ measurements has been assumed. In the example presented here, only possible construction errors according to Fig.~\ref{fig:bwire_position_variable} and Fig.~\ref{fig:pad2chip_bw_uncertain} have been considered to determine the uncertain elongations $\Delta s$ and $\Delta h$, respectively.

\begin{figure}[b]
 \centering
 \resizebox{0.8\columnwidth}{!}{\definecolor{mycolor1}{rgb}{0.00000,0.00000,0.56250}\includegraphics{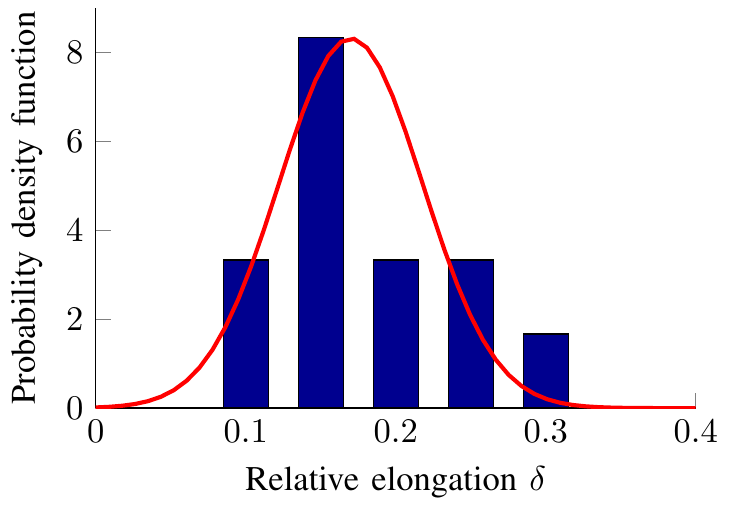}
 }
 \caption{Probability density function for the relative elongation $\delta$.}
 \label{fig:bwire_pdf}
\end{figure}

The measurement of these different length parameters has been done for one chip with $12$ bonding wire samples using the X-ray pictures shown in Fig.~\ref{fig:X-ray}. Instead of taking the total length $L$ of a bonding wire as the uncertain quantity, the relative elongation $\delta = \frac{L-d}{L}$ is used. The random elongations for all bonding wires, possibly of different lengths, are determined by the probability density function for $\delta$. From the histogram shown in Fig.~\ref{fig:bwire_pdf}, we identify a normal distribution with expectation value $\mu_{\text{BW}} = \num{0.17}$ and standard deviation $\sigma_{\text{BW}} = \num{0.048}$ albeit the rather small number of samples. A more rigorous analysis would require the fabrication of additional test chips.

\subsection{Monte Carlo Simulation}
\label{sec:monte_carlo_simulation}

Particularly, if the number of random input parameters is high, the Monte Carlo (MC) method is a well established technique to quantify the variation of outputs~\cite{Caflisch_1998aa}. To this end, MC solves repeatedly the problem, i.e., in our case (\ref{eq:electrostatics_discrete}-\ref{eq:heat_equation_discrete}), for random sets of parameters. Naturally, the error committed by taking a finite amount of samples $M$ is decreasing with increasing number of samples. It serves as a guideline for the necessary amount of samples and is approximated by
\begin{equation}
 \text{error}_{\text{MC}} = \frac{\sigma_{\text{MC}}}{\sqrt{M}},
\end{equation}
where $\sigma_{\text{MC}}$ is the standard deviation approximated with the $M$ samples. The estimator unveils the rather slow convergence in terms of $\sqrt{M}$. However, the application of other methods is straightforward.  \section{Simulation Results}
\label{sec:simulation_results}

\todo[inline]{Any simulation parameters missing in this table?}

\subsection{Chip Model}
\label{sec:chip_model}

\begin{figure}[t]
 \centering
 \subfloat[Inclined top view\label{fig:fig2_model}]{\includegraphics[height=11em]{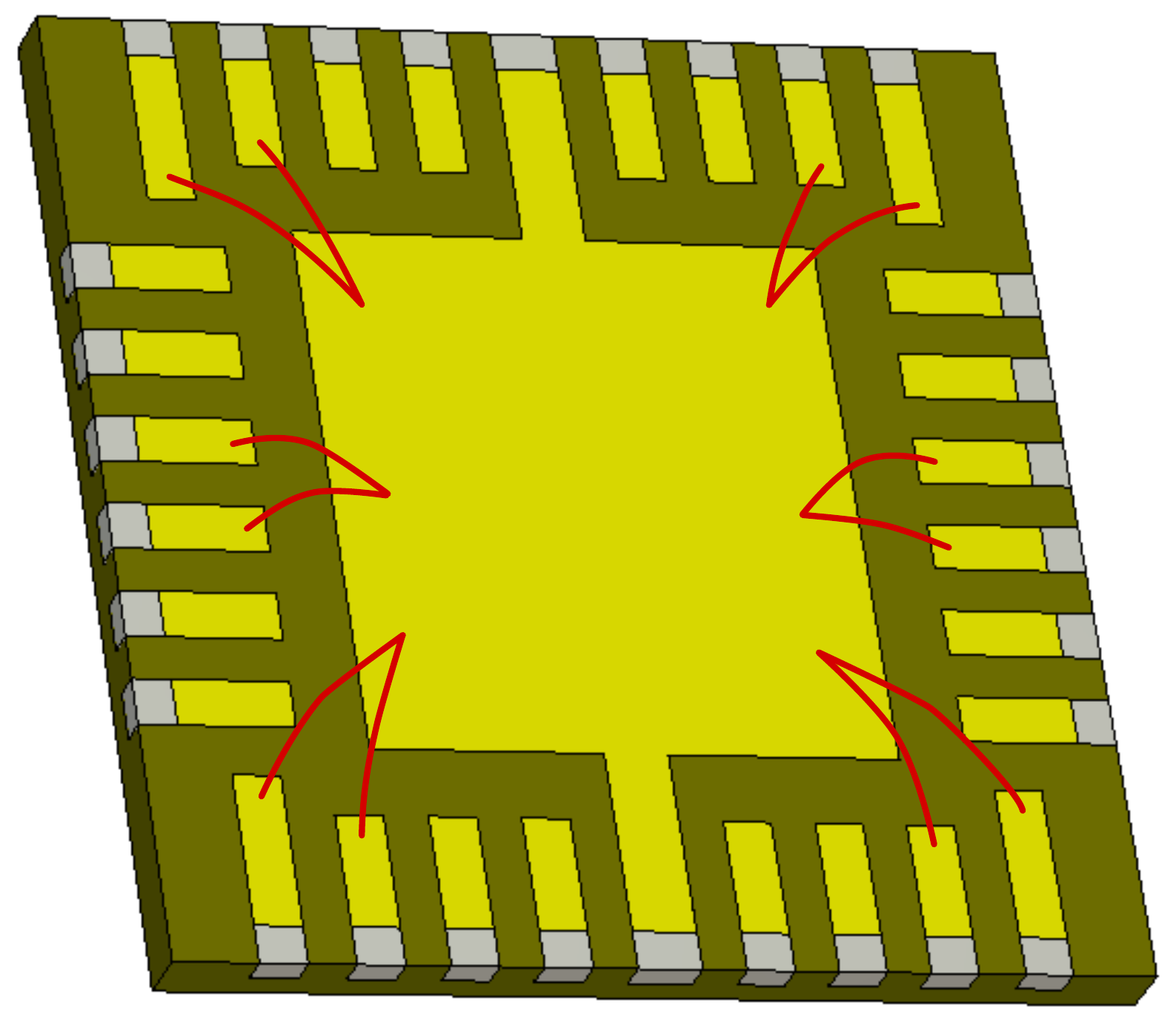}}\quad
 \subfloat[Hexahedral mesh\label{fig:fig3}]{\includegraphics[height=11em]{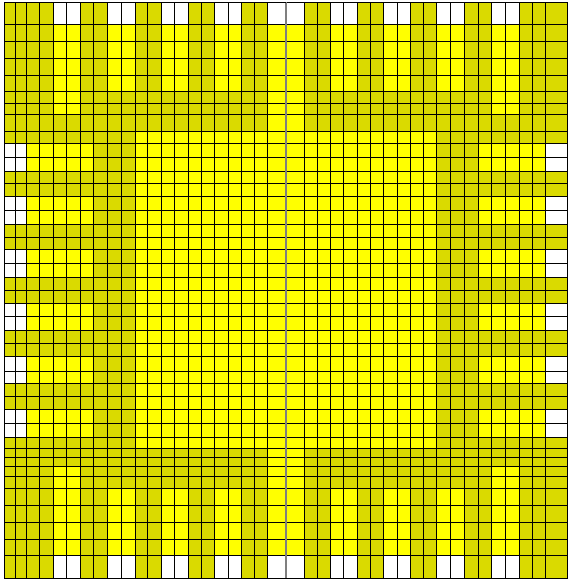}}
 \caption{Model of the investigated chip.}
 \label{fig:chip}
\end{figure}

The chip is modeled following the geometry obtained by measuring the X-ray pictures shown in Fig.~\ref{fig:X-ray}. All $28$ contact pads have been modeled to be of equal width ${w_{\text{pad}} = \SI{0.311}{mm}}$. $24$ of them have the same length ${\ell_{\text{pad}}=\SI{1.01}{mm}}$, whereas the other $4$ have a length ${L_{\text{pad}}=\SI{1.261}{mm}}$. As an approximation, all structures are approximated using rectangular shapes. Copper is chosen as the material for the bonding wires, the contact pads and the chip while epoxy resin is used for the mold compound. Additionally, the outer end of each contact pad is modeled as Perfectly Electric Conducting~(PEC). In Fig.~\ref{fig:chip}, the model and the computational mesh are shown. The corresponding materials and their conductivities at $T=\SI{300}{K}$ are collected in Table~\ref{tab:material_properties}.

\begin{table}[b]
 \centering
 \caption{Material Properties @ $T=\SI{300}{K}$}
 \label{tab:material_properties}
 \begin{tabular}{|l|c|c|c|} \hline
	Region & Material & $\lambda$ [\si{W/K/m}] & $\sigma$ [\si{S\per m}] \\ \hline
	Compound & Epoxy resin & \num{0.87} & \num{1e-6} \\
	Contact pad & Copper & \num{398} & \num{5.80e7}\\
	Chip & Copper & \num{398} & \num{5.80e7}\\
	Bonding wire & Copper & \num{398} & \num{5.80e7}\\ \hline
 \end{tabular}
\end{table}

\subsection{Boundary and Initial Conditions}

As electric boundary conditions, the PEC nodes are connected to a constant potential of ${V_{\text{dc}} = \pm\SI{20}{mV}}$ such that the voltage over each of the $6$ pairs of bonding wires equals ${V_{\text{bw}} = \SI{40}{mV}}$. For all other boundaries, current flow is prevented by setting homogeneous Neumann conditions. All non-PEC nodes are set to the initial potential ${V_{\text{init}}=\SI{0}{V}}$ at time $t=\SI{0}{s}$.

The thermal boundary conditions are as follows. For all boundaries, convection and radiation conditions with a heat transfer coefficient of $h=\SI{25}{W/m^2/K}$ and an emissivity of $\sigma_{\text{rad}}=\num{0.2475}$ are chosen, respectively. As initial conditions, the whole chip is assumed to be at the ambient temperature~${T_\infty=\SI{300}{K}}$.

\subsection{Quantities of Interest}

Since a possible failure of the bonding wires is investigated, we are interested in the temperature of the wires over time. As the wires themselves are not resolved by the grid (see section~\ref{sec:bonding_wire_model}), the representative wire temperatures are extracted from the end-points of the wires as given by \eqref{eq:wire_temperature}. The expectation value $E_j(t)$ for each wire temperature is calculated by averaging over all $M$ samples, i.e., 
\begin{equation*}
 E_j(t)=\frac{1}{M}\sum_{m=1}^{M}T_{\mathrm{bw},j}^{(m)}(t).
\end{equation*}
With the objective in mind that none of the bonding wires should fail, it is sufficient to pick out the wire that experiences the highest expected temperature $E_j(t)$. 
Therefore, we define $E_\mathrm{max}(t)$ to be the maximum of the expectation values of all wires, viz.
\begin{equation}
 E_\mathrm{max}(t) = \max_j[E_j(t)], \text{ for } j\in\{1,...,N_\mathrm{bw}\}.
\end{equation}
Other stochastic moments, for example the variance or standard deviation, can be determined analogously. 

\subsection{Discussion}

\begin{figure}[b]
 \centering
 \includegraphics{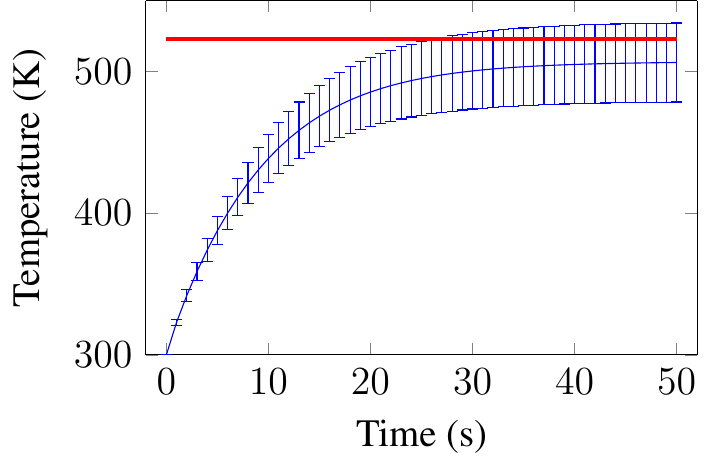}
 \caption{Expected temperature at end-point of the hottest bonding wire with plotted $6\sigma_\text{MC}$-deviation over time. In red, the critical temperature of the wire's material is indicated.}
 \label{fig:MC_output}
\end{figure}

A Monte Carlo simulation with $M = 1000$ samples was carried out using the probability density function from Fig.~\ref{fig:bwire_pdf}. With the simulation parameters as given in Table~\ref{tab:simulation_parameters}, the Monte Carlo error as introduced in section \ref{sec:monte_carlo_simulation} calculates to $\text{error}_{\text{MC}}=\SI{0.147}{K}$. In Fig.~\ref{fig:MC_output}, the resulting expectation value $E(t)$ for the temperature of the hottest wire is plotted over time. Error bars indicate the output variation resulting from the variability in the input, being the length of the wire. Assuming that a bonding wire fails mainly due to the degradation of the surrounding mold, a critical temperature $T_\text{critical} = \SI{523}{K} \approx \SI{250}{\celsius}$ is defined to mark the threshold for failure. This critical temperature is inserted as a red line to indicate the upper bound for design validity.

\begin{table}[t]
	\centering
	\caption{Simulation Parameters.}
	\label{tab:simulation_parameters}
	\begin{tabular}{|l|r|}\hline
		Parameter & Value\\ \hline
		Bonding wire voltage $V_{\text{bw}}$ & \SI{40}{mV}\\ 
		End time & \SI{50}{s}\\
		No. of time steps & \num{51} \\
		No. of MC samples & \num{1000} \\
		Wires' diameter & \SI{25.4}{\micro\metre}\\ 
		Average wires' length $\overline{\text{L}}$ & \SI{1.55}{mm}\\
		Ambient temperature & \SI{300}{K}\\
		Heat transfer coefficient & \SI{25}{W/m^2/K}\\
		Emissivity & 0.2475\\ \hline
	\end{tabular} 
\end{table}

\begin{figure}[t]
 \centering
 \includegraphics[width=\columnwidth]{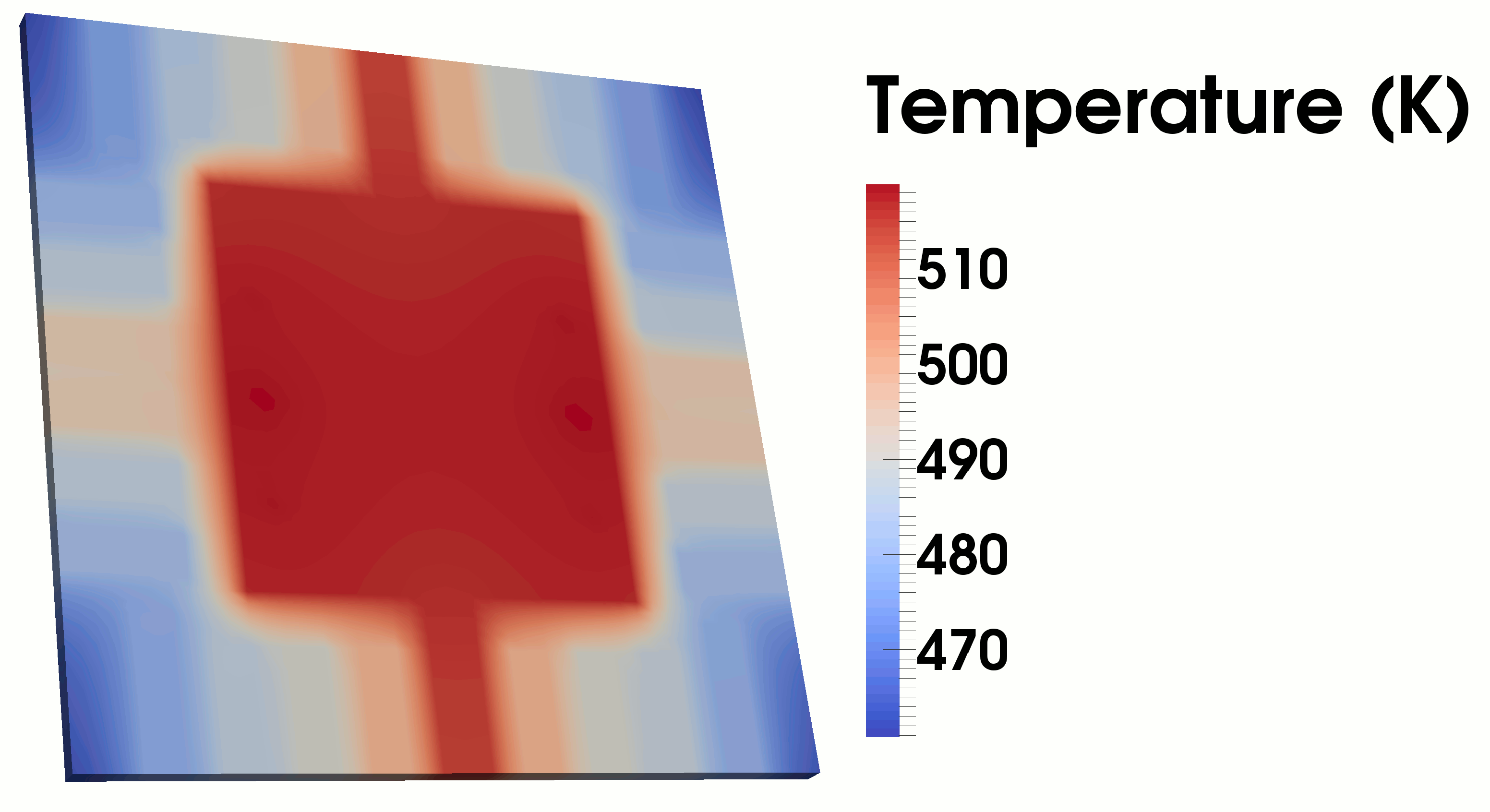}
 \caption{Spatial temperature distribution at $t=\SI{50}{s}$.}
 \label{fig:temp_distribution}
\end{figure}

Thanks to convection and radiation at the chip's boundaries, a stationary situation is observed after $t\approx\SI{50}{s}$. Then, the mean temperature of the hottest wire is still lower than the critical temperature $T_{\text{critical}}$. 
However, the uncertainty in the lengths of the bonding wires leads to variations of the temperature with a standard deviation of $\sigma_{\text{MC}}=\SI{4.65}{K}$. As the failure of bonding wires is a relevant reliability problem, the $6\sigma$-deviation is visualized by the error bars in the figure.
For the given configuration, the variation may indeed influence whether a bonding wire fails or not. This can be seen as the error bars cross the critical temperature for $t>\SI{26}{s}$. 
Fig.~\ref{fig:temp_distribution} shows the spatial temperature distribution at ${t=\SI{50}{s}}$. As one would expect, the region where the contacts are closest and are connected by the shortest wires experience the largest temperature increase. These wires are the most sensitive in the system. This is confirmed by the fact that one of these wires is the one with the maximal temperature evolution shown in Fig.~\ref{fig:MC_output}.

 \section{Conclusions}
\label{sec:conclusions}

Bonding wires are included as lumped elements in a coupled electrothermal field model. By this, the behavior of the bonding wires can be checked already in the design phase. The thickness and material of the bonding wires can be selected by evaluating the simulation results. 

As a possible application, a stochastic model for uncertain bonding wire lengths has been set up for an exemplary chip package. This uncertainty is included in the coupled electrothermal field model and affects the final operating temperature of the bonding wires. It is quantified in terms of expectation value and standard deviation.
Concerning the investigated uncertainty of the bonding wires' lengths, it is important to recall that the used data set to extract the probability density function is very small. However, the simulation result of the presented example indicates that the relative influence of the uncertainties can be significant for the validity of bonding wire design.

Future research will incorporate more sophisticated bonding wire models and a comparison to bonding wire measurements.

\section*{Acknowledgment}
The authors would like to thank Jiri Petrzela, Roman Sotner and Jiri Drinovsky for the bonding wire measurements done at Brno University of Technology.


\begin{thebibliography}{1}

\bibitem{Clemens_2005aa} M.~Clemens, \emph{Large Systems of Equations in a Discrete Electromagnetism: Formulations and Numerical Algorithms}, IEE Proceedings - Science, Measurement and Technology, 152(2), March 2005.

\bibitem{Clemens_2001ac}
M.~Clemens, E.~Gjonaj, P.~Pinder, and T.~Weiland.
\newblock Self-Consistent Simulations of Transient Heating Effects in
  Electrical Devices Using the Finite Integration Technique.
\newblock {\em IEEE Transactions on Magnetics}, 37(5):3375--3379, September
  2001.

\bibitem{Duque_2015aa}
D.~J.~Duque, S.~Schöps, and A.~Wieers.
\newblock Fast and Reliable Simulations of the Heating of Bond Wires.
\newblock In Giuseppe Nicosia, editor, {\em Progress in Industrial Mathematics
  at {ECMI} 2014}, Mathematics in Industry, Berlin, 2015. Springer. Accepted for publication.

\bibitem{Alotto_2010aa}
P.~Alotto, F.~Freschi und M.~Repetto.
\newblock Multiphysics Problems via the Cell Method: The Role of Tonti Diagrams.
\newblock {\em IEEE Transactions on Magnetics}, 46(8):2959–2962, 2010. 

\bibitem{Mouthaan_2001aa}
K.~Mouthaan.
\newblock {\em Modelling of RF High Power Bipolar Transistors}.
\newblock PhD thesis, TU Delft, Delft, The Netherlands, 2001.

\bibitem{Nobauer_2000aa}
G.~T. Nöbauer and H.~Moser.
\newblock Analytical Approach to Temperature Evaluation in Bonding Wires and
  Calculation of Allowable Current.
\newblock {\em IEEE Transactions on Advanced Packaging}, 23(3):426---435,
  August 2000.

\bibitem{Schoenmaker_2013aa}
W.~Schoenmaker et al.
\newblock Fully-Coupled 3D Electro-Thermal Field Simulator for Chip-Level
  Analysis of Power Devices.
\newblock In {\em 19th International Workshop on Thermal Investigations of ICs
  and Systems (THERMINIC)}, pages 210--215, Berlin, Germany, 2013. IEEE.

\bibitem{Schuster_2000aa}
C.~Schuster, G.~Leonhardt and W.~Fichtner.
\newblock Electromagnetic Simulation of Bonding Wires and Comparison with Wide Band Measurements.
\newblock {\em IEEE Transactions on Advanced Packaging}, 23(1):69--79, February
  2000.

\bibitem{Weiland_1977aa}
T.~Weiland.
\newblock A Discretization Model for the Solution of {Maxwell}'s Equations for Six-Component Fields.
\newblock {\em International Journal of Electronics and Communications},
  31:116--120, March 1977.

\bibitem{Weiland_1996aa}
T.~Weiland.
\newblock Time Domain Electromagnetic Field Computation with Finite Difference Methods.
\newblock {\em International Journal of Numerical Modelling: Electronic
  Networks, Devices and Fields}, 9(4):295--319, 1996.

\bibitem{Caflisch_1998aa}
R.~E. Caflisch.
\newblock Monte Carlo and Quasi-Monte Carlo Methods.
\newblock {\em Acta Numerica}, 7:1--49, 1998.

\end{thebibliography}
\end{document}